# Intrinsic frequency distribution characterises neural dynamics

**Authors:**

Ryohei Fukuma[a,b], Yoshinobu Kawahara[c,d], Okito Yamashita[c,e], Kei Majima[f], Haruhiko Kishima[b], Takufumi Yanagisawa[a,b*]

**Affiliations:**

[a] Department of Neuroinformatics, Graduate School of Medicine, University of Osaka, 2-2 Yamadaoka, Suita, Osaka 565-0871, Japan

[b] Department of Neurosurgery, Graduate School of Medicine, University of Osaka, 2-2 Yamadaoka, Suita, Osaka 565-0871, Japan

[c] RIKEN Center for Advanced Intelligence Project, 15th floor, Nihonbashi 1-chome Mitsui Building, 1-4-1 Nihonbashi, Chuo-ku, Tokyo 103-0027, Japan

[d] Department of Information Systems Engineering, Graduate School of Information Science and Technology, University of Osaka, 1-5 Yamadaoka, Suita, Osaka 565-0871 Japan

[e] Department of Computational Brain Imaging, ATR Neural Information Analysis Laboratories, 2-2-2 Hikaridai, Seika-cho, Kyoto 619-0288, Japan

[f] Institute for Quantum Life Science, National Institutes for Quantum Science and Technology, 4-9-1 Anagawa, Inage-ku, Chiba, Chiba 263-8555, Japan

**\*Corresponding author:** Takufumi Yanagisawa, M.D.-Ph.D.

Department of Neuroinformatics, Graduate School of Medicine, University of Osaka, 2-2 Yamadaoka, Suita, Osaka 565-0871, Japan

Tel.: +81-66-879-3652

Fax: +81-66-879-3659

E-mail: tyanagisawa@nsurg.med.osaka-u.ac.jp

**Abstract**

Decomposing multivariate time series with certain basic dynamics is crucial for understanding, predicting and controlling nonlinear spatiotemporally dynamic systems such as the brain. Dynamic mode decomposition (DMD) is a method for decomposing nonlinear spatiotemporal dynamics into several basic dynamics (dynamic modes; DMs) with intrinsic frequencies and decay rates. In particular, unlike Fourier transform-based methods, which are used to decompose a single-channel signal into the amplitudes of sinusoidal waves with discrete frequencies at a regular interval, DMD can derive the intrinsic frequencies of a multichannel signal on the basis of the available data; furthermore, it can capture nonstationary components such as alternations between states with different intrinsic frequencies. Here, we propose the use of the distribution of intrinsic frequencies derived from DMDs (DM frequencies) to characterise neural activities. The distributions of DM frequencies in the electroencephalograms of healthy subjects and patients with dementia or Parkinson's disease in a resting state were evaluated. By using the distributions, these patients were distinguished from healthy subjects with significantly greater accuracy than when using amplitude spectra derived by discrete Fourier transform. This finding suggests that the distribution of DM frequencies exhibits distinct behaviour from amplitude spectra, and therefore, the distribution may serve as a new biomarker by characterising the nonlinear spatiotemporal dynamics of electrophysiological signals.

**Introduction**

Spatiotemporal dynamics such as neural activities have been characterised by a combination of oscillations with multiple intrinsic frequencies and amplitudes in phase relations[1]. The intrinsic frequency ranges from 0.05 to 500 Hz at different scales from single neurons in microcircuits to neural masses at the macroscopic level of the mammalian brain[2,3]. These oscillatory dynamics significantly vary depending on the animal's behaviour and neurological state. In the development of electroencephalography (EEG), Hans Berger discovered that there are two intrinsic frequencies of EEG signals, alpha and beta waves, and that the amplitude of alpha waves is significantly decreased by eye opening[4]; that is, the amplitude of alpha waves represents the neural states of eye opening and closing. The intrinsic frequencies of EEG signals for a single subject also vary depending on the state of neural activity, such as sleep-wake[5] cycle and cognitive status[6,7]; moreover, intrinsic frequencies vary among subjects depending on the presence of diseases such as epilepsy[8], chronic pain[9-11], dementia[12], and Parkinson's disease (PD)[13]. These changes in the oscillatory dynamics of multiple intrinsic frequencies can be used to characterise neural activity, but clearly decomposing changes in the EEG waveform into changes in the intrinsic frequencies and their respective amplitudes is difficult.

Fourier transform-based signal decomposition, specifically discrete Fourier transform (DFT), has long been used as a model-free method for characterising the dynamics of oscillatory systems with various frequencies[14]; however, DFT may not be optimal for assessing spatiotemporal signals that are composed of multiple nonstationary oscillatory signals, such as neural activity[2]. DFT can be used to decompose a signal into stationary oscillations with discrete frequencies that do not necessarily match their intrinsic frequencies. The frequency resolution is determined on the basis of the duration and sampling rate of the signal; hence, DFT cannot be used to distinguish differences in oscillations that are smaller than the frequency resolution. In addition, because the stationarity of the signal within the given time window is assumed in DFT, this approach is not good for cases with decay or divergence of the oscillatory component, e.g., signals with changing states, such as on/off. Furthermore, it is difficult for DFT to handle multichannel signals via relationships among channels. Therefore, to assess the intrinsic frequencies of neural activities, suitable signal decomposition methods must meet the following requirements: (1) handle multichannel (i.e., spatiotemporal) signals and (2) decompose signals into intrinsic frequencies and their respective amplitudes with decay or growth over time.

Dynamic mode decomposition (DMD) is a data-driven, model-free method that can be used to reduce the dimensionality of high-dimensional time series by extracting a low-dimensional basis, called dynamic modes (DMs), which approximate Koopman modes[15-20]. Specifically, DMD is used to extract the main components of a multichannel signal by performing singular value decomposition (SVD); using these extracted components, the original signal is represented as the sum of multiple spatial patterns—DMs—with corresponding intrinsic frequencies and amplitudes that decay or diverge over time. DMD was proposed by Schmid *et al*. in 2008[21] and has been applied to various high-dimensional time series of data, such as fluid dynamics data[15,17,20], satellite observations[22], infectious disease data[23], and sleep spindles recorded via electrocorticographic (ECoG) signals[24], demonstrating its effectiveness. The characteristics derived from DMD can be used to successfully classify functional magnetic resonance imaging scans of subjects with different traits[25] and of patients with major depressive disorders[26] and further classify ECoG signals corresponding to certain hand movements and visual stimuli[27,28]. Notably, the accuracy of classification varies depending on the number of components extracted via SVD[28], suggesting that SVD approximation, which is based on SVD component selection, contributes to the extraction of informative features. Moreover, DMD is associated with phase reduction, whereby the dynamical equations of an oscillatory system are simplified to a one-dimensional phase equation. While phase reduction has been used as a dynamical model to compare synthetic data with actual observations, DMD can estimate limit cycles with frequencies that match the dynamics of the observed data[24,29]. Therefore, the intrinsic frequencies estimated by DMD using appropriate SVD components differ from the frequencies that form the basis of frequency decomposition in DFT, and the frequencies themselves correspond to the dynamics of high-dimensional time series data and characterise those dynamics.

In this study, we hypothesised that the distribution of intrinsic frequencies estimated by DMD (1) serves as a feature that is distinct from amplitude spectra acquired with DFT and (2) better characterises nonstationary high-dimensional time series, such as neural activity series. In other words, DMD is employed to separate the variations in intrinsic frequency and amplitude, and the contribution of estimated intrinsic frequencies to the classification of spatiotemporal dynamics is evaluated. We applied DMD to two resting-state EEG datasets for patients with dementia and PD, in order to empirically evaluate the effectiveness of estimated intrinsic frequencies to classify diseases; these results were then compared with those obtained via the amplitude spectra and spatial DM (sDM) methods proposed in a previous study[28].

**Results**

To evaluate the ability of DMD to capture intrinsic frequencies, a simple signal source is considered. This signal source generates 10 channels of signals ($x_i(t)$, where $i$ and $t$ denote the channel ($i = 1, \cdots, 10$) and time [s], respectively) at a sampling rate of 500 Hz, and of these, 2 channels (channels 1 and 2) output 10 and 20 Hz sinusoidal waves, in addition to Gaussian noise ($N(0,1)$) with $\sigma = 2.5$ for all channels. Both the 10 and 20 Hz sinusoidal waves have an amplitude of 1 and are always present (stationary condition).

Stationary condition:
$$\begin{cases} x_i(t) := 1 \cdot \sin(2\pi \cdot 10 \cdot t + \varepsilon_p) + 1 \cdot \sin(2\pi \cdot 20 \cdot t + \varepsilon_p) + 2.5 \cdot N(0,1) \text{ where } i \in \{1,2\} \\ x_i(t) := 2.5 \cdot N(0,1) \text{ where } i \in \{3,\cdots,10\} \end{cases}$$
where $\varepsilon_p \sim U(-\pi, \pi)$, with $U$ denoting a continuous uniform distribution.

Let us consider the decomposition of one second of this signal via DMD (Fig. 1a). First, a signal ($\mathbf{x}(t)^T = [x_1(t)\ x_2(t) \ldots x_P(t)] \in \mathbb{R}^P$; $P$ denotes the number of observation points, and $P = 10$ in this case) sampled with an interval of $\Delta t$ (which is 2 ms, in this case) is decomposed via SVD to obtain 454 components (Fig. 1b; see signal stacking in the Methods). On the basis of these SVD components, DMD is used to approximately decompose $\mathbf{x}(t)$ as a superposition of $K$ oscillatory components in a complex space, as follows, by determining $\lambda_k$ and $\boldsymbol{\varphi}_k$:

$$\mathbf{x}(t) \approx \sum_{k=1}^{K} \boldsymbol{\varphi}_k\ r_k^t \exp(2\pi i\ f_k\ t)\ b_k$$

where $r_k = |\lambda_k|^{1/\Delta t}$ and $f_k = \arg(\lambda_k)/2\pi\Delta t$.

Here, each of the $K$ oscillatory components is represented with the following parameters: $\boldsymbol{\varphi}_k$, the DM denoting the spatial pattern ($P$-dimensional complex vector); $f_k$, the frequency of the DM; $r_k$, the decay/growth rate of the DM; and $b_k$, the initial phase of the DM (complex scalar). For the decomposition of this signal, DMs are obtained on the basis of the first two components because these two components have particularly large singular values (red shaded area in Fig. 1b); consequently, two DMs

($\boldsymbol{\varphi}_k$) and the corresponding $\lambda_k$ are obtained (Fig. 1c). Note that the number of DMs obtained is the same as the number of SVD components used (in this case, $K = 2$). These two DMs (and $\lambda$) are complex conjugates that represent a single-component signal in the real number space. The DMs obtained here are regularised via L2 regularisation, and the following sDM features are obtained by using a matrix ($\boldsymbol{\phi}$) with the regularised DMs as columns (Fig. 1d).

$$\text{sDM features} := \boldsymbol{\phi}\boldsymbol{\phi}', \text{ where } \boldsymbol{\phi} := \left[\frac{\boldsymbol{\varphi}_1}{\|\boldsymbol{\varphi}_1\|_2} \cdots \frac{\boldsymbol{\varphi}_K}{\|\boldsymbol{\varphi}_K\|_2}\right]$$

On the other hand, the frequency of the two DMs (DM frequencies) is 20.5 Hz (Fig. 1e). Notably, the DM frequencies become 9.7 and 20.5 Hz (Fig. 1f) when the first four SVD components of the same signal are used to obtain DMs ($K = 4$, red and pink shaded areas in Fig. 1b); these two frequencies correspond to two peaks at approximately 10 and 20 Hz in the amplitude spectrum of the same signal evaluated by DFT (Fig. 1g).

Similarly, one second of the signal ($x_i(t)$) was generated 100 times (trials), changing the random seed used to generate $\varepsilon_p$. DMD was applied to the signal of each trial to obtain DMs with $K = 2$. Notably, the intrinsic frequencies estimated with DMD were distributed near 10 or 20 Hz, depending on the trial (Supplementary Fig. 1). In all 100 trials, the number of mode occurrences corresponding to each 1 Hz frequency bin for the DM frequencies was counted. The distribution of the DM frequencies showed comparable peaks at 10 and 20 Hz (Fig. 1h, green), which were similar to those observed in the amplitude spectrum averaged across 100 trials (Fig. 1i, green). Although intrinsic frequencies were estimated using only the first two SVD components, two signals with the same amplitude but different frequencies were detected with similar probabilities in 100 trials.

Next, let us consider a nonstationary condition in which a 10 Hz sinusoidal wave is intermittently output and a 20 Hz sinusoidal wave continues to be output stationarily. Here, the 20 Hz sinusoidal wave has an amplitude of 1 and is always present, and the 10 Hz sinusoidal wave has an amplitude of 2 and the duration of the wave is randomly determined between 10% and 90% of the 1 s time window (blue lines in Supplementary Fig. 1a), which is controlled by $f(t) \in \{0, 1\}$ in the following equation.

Nonstationary condition:
$$\begin{cases} x_i(t) := 2 \cdot f(t) \cdot \sin(2\pi \cdot 10 \cdot t + \varepsilon_p) + 1 \cdot \sin(2\pi \cdot 20 \cdot t + \varepsilon_p) + 2.5 \cdot N(0,1) \text{ where } i \in \{1,2\} \\ x_i(t) := 2.5 \cdot N(0,1) \text{ where } i \in \{3, \cdots, 10\} \end{cases}$$

where $f(t) = (1 + sign(\sin(\pi t + \varepsilon_t)))/2$ with $\varepsilon_t \sim U(-0.9\pi, -0.1\pi)$ or $U(0.1\pi, 0.9\pi)$.

The signals were again generated 100 times (trials), with the random seed changed to generate $\varepsilon_p$ and $\varepsilon_t$. Notably, $\varepsilon_t$ was controlled so that the duration of the 10 Hz sinusoidal wave was 50% of the time window on average over 100 trials. In other words, the 10 Hz sinusoidal wave had an average amplitude of 1 over 100 trials. Interestingly, when these signals were converted to DMs via the first two SVD components, there were more cases in which the DM frequency was 10 Hz (intermittent with a larger amplitude) than 20 Hz (stationary with a smaller amplitude). Consequently, the DM frequencies displayed different distributions between the nonstationary condition (Fig. 1h, blue) and stationary condition (Fig. 1h, green). On the other hand, the amplitude spectra under nonstationary and stationary conditions were similar (Fig. 1i). In addition, obtaining DMs from the first four SVD components resulted in similar distributions of the DM frequencies (Supplementary Fig. 1f). Therefore, it was suggested that the distribution of DM frequencies, i.e., the distribution of intrinsic frequencies estimated with DMD, can be used to distinguish stationary and nonstationary oscillations, which are difficult to distinguish in the amplitude spectra obtained with DFT, with an appropriate number of SVD components.

Finally, considering that the sDM features exhibited similar patterns between the two conditions (Fig. 1j), these results demonstrate that the distribution of the DM frequencies—the intrinsic frequencies estimated with DMD from multichannel signals—can be used to characterise the differences in the stabilities of the intrinsic oscillations in the signal source, which are difficult to assess with amplitude spectra or sDM features.

***Behaviours of DM frequencies and amplitude spectra for patients with dementia***
To assess the distribution of DM frequencies as a biomarker of neurological diseases, the proposed method was applied to characterise healthy subjects and patients with dementia due to Alzheimer's disease (AD) and frontotemporal dementia (FTD) via EEG signals during eyes-closed resting states[30]. The dataset consisted of resting-state EEG signals from 29 age-matched healthy subjects, 36 patients with AD, and 23 patients with FTD (Methods). The duration of the signals for each subject was approximately 13.2 min (5.1-21.3 min).

First, the DM frequencies from the resting-state EEG signals were compared with the amplitude spectra derived from the same EEG signals via DFT. The DMs were calculated for the EEG signals divided over nonoverlapping 1 s time windows. The time series of the DM frequencies for representative subjects are shown in Fig. 2a. For the healthy subject, the DM frequencies were distributed narrowly at approximately 10 Hz (Fig. 2a top row; number of SVD components: 50), resulting in a sharp peak in the distribution of the DM frequencies at approximately 10 Hz (Fig. 2b, top row). The amplitude spectrum also showed a clear peak at 10 Hz (Fig. 2c, top row). On the other hand, for the patient with AD, the DM frequencies were distributed broadly below 10 Hz (Fig. 2a, middle row), resulting in a lower frequency of the peak (Fig. 2b, middle row); moreover, the amplitude spectrum did not show a clear peak (Fig. 2c, middle row). For the patient with FTD, the peak occurred at 10 Hz (Fig. 2a, b, bottom row), similar to that of the healthy subject; however, DMs at approximately 6 Hz also increased in number compared with those of the healthy subject. The amplitude spectrum clearly peaked at 10 Hz (Fig. 2c, bottom row).

***Classification of patients with dementia and healthy subjects according to the distribution of DM frequencies***

To evaluate the difference in amplitude spectra among each participant group, the amplitude spectra of each subject were averaged first among the time windows and then among the subjects in each group. All three groups had a prominent peak in the alpha band, and the peak amplitude significantly differed between healthy subjects and patients with dementia (FTD > AD) (Bonferroni-adjusted $p = 0.0091$, one-way ANOVA; Fig. 3a). In addition, there were significant differences in amplitude in the beta band. However, no significant differences were observed in the other frequency bands. In the topographic map of the peak frequency of the alpha band (9.8 Hz for healthy subjects, indicated by the arrow in Fig. 3a), the alpha amplitude in the occipital region differed between the patients with dementia and healthy subjects (Fig. 3b), which was consistent with the findings of previous studies[31,32].

On the other hand, when DMD was applied to the same EEG signals and time windows, the distributions of the DM frequencies were significantly different among the three participant groups in the theta, alpha, and beta bands (Bonferroni-adjusted $p < 0.05$, one-way ANOVA; number of SVD components: 50; Fig. 3c). Similar to the amplitude spectrum, healthy subjects presented higher values than patients with dementia did in the alpha band. A similar trend was also observed in the beta band. On the other hand, in the theta band, patients with dementia presented higher values than healthy subjects did. This increase was particularly prominent in patients with AD.

Figure 3d illustrates the sDM features averaged among each group. The spatial node DM features (diagonal components of the matrix; snDM features) presented high values corresponding to the electrodes in the occipital and temporal regions, as did the alpha amplitude. In addition, the spatial edge DM features (nondiagonal components of the matrix; seDM features) between these electrode pairs also presented high values. These snDM and seDM features differed among the three groups and presented high $F$ values (Fig. 2e). In other words, similar to amplitude, the sDM features related to the electrodes in the occipital and temporal regions differed between patients with dementia and age-matched healthy subjects.

Here, three groups of subjects were classified based on either amplitude features or temporal frequency DM (tfDM) features via a linear support vector machine (SVM). To optimise the hyperparameters (cost parameter for the classifier and number of SVD components used in DMD), 10-fold nested cross-validation was employed. The amplitude features and tfDM features were obtained as the averages of the amplitude spectra and number of mode occurrences within conventional frequency bands (low-delta: 0–1 Hz, high-delta: 1–4 Hz, theta: 4–8 Hz, alpha: 8–13 Hz, beta: 13–30 Hz, low-gamma: 30–80 Hz, high-gamma: 80–250 Hz). The classification accuracy was significantly highest when tfDM features were used ($p < 0.001$, $F(4, 45) = 17.48$ one-way ANOVA; amplitude: 56.96 ± 2.05% (mean ± 95% confidence interval among ten repetitions of classification), snDM: 55.07 ± 3.39%, seDM: 54.49 ± 2.65%, tfDM: 65.49 ± 1.61%, sDM + tfDM: 55.29 ± 2.33%; Fig. 3f; for classification accuracies versus the number of SVD components, see Supplementary Fig. 2a). Notably, the classification accuracy with tfDM features was significantly higher than that with amplitude features ($p < 0.001$, post hoc Tukey–Kramer test). In addition, the confusion matrix revealed that while the amplitude and sDM features often led to the misclassification of patients with FTD as having AD, the tfDM feature reduced the number of misclassified patients (Fig. 3g). These results suggest that the proposed tfDM features can be used to more accurately distinguish between the resting-state EEG signals of healthy subjects and patients with dementia than when DFT-based features are used.

***Classification of patients with PD and healthy subjects***

Using the proposed method, the classification of resting-state EEG signals for patients with PD and healthy subjects[33] was performed. The analysis was performed by comparing the EEG signals of healthy subjects with their eyes closed and the EEG signals of patients on medication with their eyes closed. The number of both healthy subjects and patients was 27, and the measurement duration was 1 minute.

Although there was no significant difference (Bonferroni-adjusted $p \geq 0.05$, one-way ANOVA; Fig. 4a), the resting-state EEG signals of patients with PD showed an increase in amplitude across a wide frequency band and cortical area (Fig. 4a, b). The distribution of the DM frequencies revealed an increase in the low-frequency band (< 4 Hz) of patients with PD (number of SVD components: 20; Fig. 4c), although the difference was not significant (Bonferroni-adjusted $p \geq 0.05$, one-way ANOVA), except at 0–1 Hz (not shown in Fig. 4c). Notably, the peak in the alpha band decreased in frequency (healthy subjects: 12–13 Hz, patients with PD: 8–9 Hz). Neither the snDM nor seDM features presented large values, and the differences between the participant groups were small (Fig. 4d, e; for all features corresponding to all electrodes, see Supplementary Fig. 3). When the classification accuracies between these two groups using different features were compared, the highest accuracy was achieved with the tfDM features ($p < 0.001$, $F(4, 45) = 19.56$ one-way ANOVA with post hoc Tukey–Kramer test; amplitude: 67.22 ± 2.65%, snDM: 60.19 ± 4.15%, seDM: 60.93 ± 3.56%, tfDM: 73.70 ± 2.85%, sDM + tfDM: 71.67 ± 1.98%; Fig. 4f, g; for classification versus against the number of SVD components, see Supplementary Fig. 2b). Notably, the use of tfDM features significantly increased the classification accuracy compared with the use of amplitude features ($p = 0.015$, post hoc Tukey–Kramer test). Overall, the resting-state EEG signals of patients with PD are characterised by some peak shifts in the intrinsic frequencies in the alpha and beta bands.

**Discussion**

We propose that the distribution of DM frequencies effectively characterises high-dimensional time series data independent of sDM or DFT-based features (e.g., amplitude spectrum). In the simulation, it was shown that the distribution of the DM frequencies captures differences in the stabilities of multichannel signals with two intrinsic frequencies that cannot be captured by DFT. Furthermore, it was shown that the tfDM features exhibit distinct behaviour from the amplitude features extracted by DFT from actual resting-state EEG signals. In particular, when classifying the resting-state EEG signals of patients with dementia or PD, the accuracy was significantly better when tfDM features were used than when amplitude features obtained via DFT were used. In other words, distribution of DM frequencies contains information that cannot be captured by DFT, allowing tfDM features to be independent of both amplitude and sDM features. These results suggest that DMD was able to extract a representation that accurately reflected oscillatory dynamics by separating the intrinsic frequency and corresponding amplitude of each oscillatory activity, which had previously been expressed only as increases or decreases in amplitude spectrum. By extracting the distribution of intrinsic frequencies via DMD, it was possible to characterise oscillatory states that were previously unclear.

Our results demonstrate that the intrinsic frequency assessed on the basis of DMD can be used a biomarker for dementia and PD. Previous studies have shown that the frequency characteristics of time series data are biomarkers of various neurological diseases, although they were not separated in terms of frequency and amplitude. For example, the amplitudes of various frequency ranges in neural oscillations are biomarkers of PD[34], dementia[35], and autism spectrum disorder[36]. In other words, the amplitude spectrum varies depending on the disease. Furthermore, a peak in the alpha band is a characteristic of the EEG signals of healthy subjects[11], and the frequency of the peak varies depending on the cortical site[37], as does the cognitive status[6,7]. However, a shift in peak frequency occurs in patients with chronic pain[9,10], dementia[12], and PD[13], and this shift can be used as a biomarker. These changes in the oscillatory dynamics of multiple intrinsic frequencies are characteristic of neural activity, but it is difficult to fully capture changes in intrinsic frequencies using amplitude spectra alone. In fact, when tfDM features derived from the distribution of the DM frequencies were used, the classification accuracy of healthy subjects and patients with dementia was significantly greater than that when amplitude features were used; this was also the case for classifying healthy subjects and PD patients. In other words, by using DMD to characterise the amplitude and intrinsic frequency separately, it was possible to obtain new biomarkers other than amplitude spectra.

In patients with FTD, there was little difference in alpha amplitude compared with that of healthy subjects, but there was a large difference in the distribution of the DM frequencies in the theta and alpha bands; this difference may be related to the microstates of cortical activity or the stability of the intrinsic frequencies. Cortical activity in the whole brain measured via EEG is known to have quasi-stable "microstates" that switch in a short time of 40–120 ms[38]. The state transition and duration of the state vary for various diseases, such as dementia[39], depression[40], schizophrenia[39], and PD[41]. In the simulations in this study, even short-duration activities (intermittent 10 Hz signals in Fig. 1) within the analysis time window affected the distribution of the DM frequencies (Fig. 1h; see Supplementary Fig. 1). Similarly, the transitions between microstates that switch over a short period may even generate differences in the distribution of the DM frequencies, resulting in improved classification accuracy for participant groups. The distribution of the DM frequencies can potentially be used to characterise disease-related differences in EEG signals that vary with the instability of oscillatory states, even for oscillations with the same intrinsic frequencies.

In this study, we demonstrated that (1) DMD can be used to assess the intrinsic frequency of high-dimensional time series on the basis of the available data, and (2) the obtained distribution of DM frequencies more accurately characterises the spatiotemporal dynamics of neural activity than does that obtained with DFT, thereby serving as an independent information source and a new biomarker to identify diseases. Although there are similar data-driven methods for obtaining the frequencies of time series of data, DMD has several advantages over other methods for assessing intrinsic frequencies[42]. While DFT is a conventional method for assessing the frequency components of time series data, it assumes stationarity of signal and may therefore be unsuitable for assessing oscillatory signals that decay or diverge. In addition, since DFT can only handle single-channel signals, it is difficult to use DFT to assess the dynamics of coupled oscillators as an entire system. On the other hand, principal component analysis (PCA), proper orthogonal decomposition (POD) and empirical mode decomposition (EMD) are popular methods for assessing the dynamics of high-dimensional data through dimension reduction. Although these methods can be used for dimensionality reduction for static signals, they are not suitable for extracting dynamic information. PCA and POD ignore temporal information in signals and therefore cannot capture dynamic information[29]. This contrasts with DMD, which decomposes signals via sinusoidal waves as basis functions, making it easier to understand the physical meaning of the obtained components. On the other hand, EMD can also utilise temporal information; however, the modes obtained on the basis of data (called intrinsic mode functions) are generally nonsinusoidal waveforms that are difficult to interpret[43]. Therefore, it is difficult to accurately assess the dynamics of coupled oscillatory systems, especially the intrinsic frequencies of the systems, with these methods.

Compared with other modal decomposition methods, DMD is a reliable method for representing the intrinsic frequencies of oscillators.

**Conclusion**

The distribution of the DM frequency better characterises nonstationary spatiotemporal dynamics, such as neural activity, than do the results of DFT; furthermore, this distribution is an effective new biomarker for identifying neural diseases.

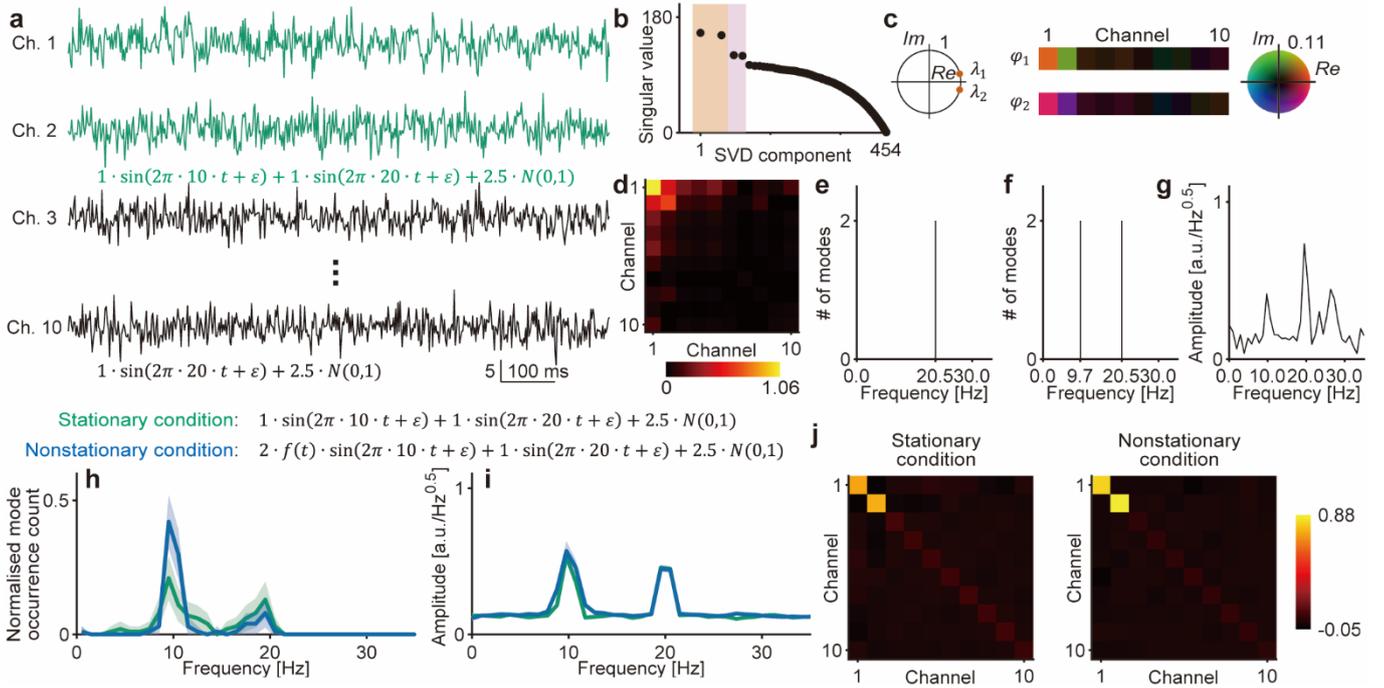

**Fig. 1. Behaviour of the DM frequencies and amplitude spectra for stationary and nonstationary signals.** (**a**) A representative trial of the spatiotemporal signal for the stationary condition over one second is visualised. Each trial consists of 10 channels, and the sampling rate of the signal is 500 Hz. The first two channels contain sinusoidal waves of 10 and 20 Hz, both with amplitudes of 1 (green lines). The initial phase of the sinusoidal waves ($\varepsilon$) was determined randomly for each trial and channel. Gaussian noise ($\sigma = 2.5$) was added to all channels at all sampling points (see Methods). (**b**) The singular values obtained by applying singular value decomposition (SVD) to the signals in **a** are shown, starting with the largest values. Following signal stacking, 454 SVD components were obtained (see Methods). The horizontal axis is on a logarithmic scale. (**c**) The dynamic modes (DMs; $\varphi_k$) and corresponding $\lambda_k$ obtained via the first two SVD components in the red shaded area of **b** are visualised. $\lambda_k$ encompasses the frequency and the rate of decay/divergence of the corresponding DM. (**d**) The spatial DM (sDM) features obtained from the two DMs in **c** are visualised. These features are represented as the matrix $P \times P$, where $P$ is the number of channels. (**e**) The frequency of DMs (DM frequency) obtained on the basis of the two $\lambda$ values in **c** is visualised. The two DMs had a frequency of 20.5 Hz. (**f**) The DM frequencies acquired via the first four SVD components (red and pink shaded areas in **b**) are visualised. The four DMs had frequencies of 9.7 and 20.5 Hz. (**g**) Amplitude spectrum acquired based on the signals from channel 1 (shown in **a**), as also analysed in **b** to **f**, is visualised. (**h-j**) For the 100 trials of stationary condition signals (shown in **a**) and nonstationary

condition signals, the (**h**) mean distributions of the DM frequencies, (**i**) mean amplitude spectra for channel 1, and (**j**) mean sDM features are shown. The shaded areas in **h** and **i** denote the 95% confidence intervals (CIs) among the trials. In the nonstationary condition, the signals of channels 1 and 2 consisted of a 10 Hz sinusoidal wave with an amplitude of 2 for approximately 50% of the duration, in addition to a continuous 20 Hz sinusoidal wave with an amplitude of 1 (see Methods).

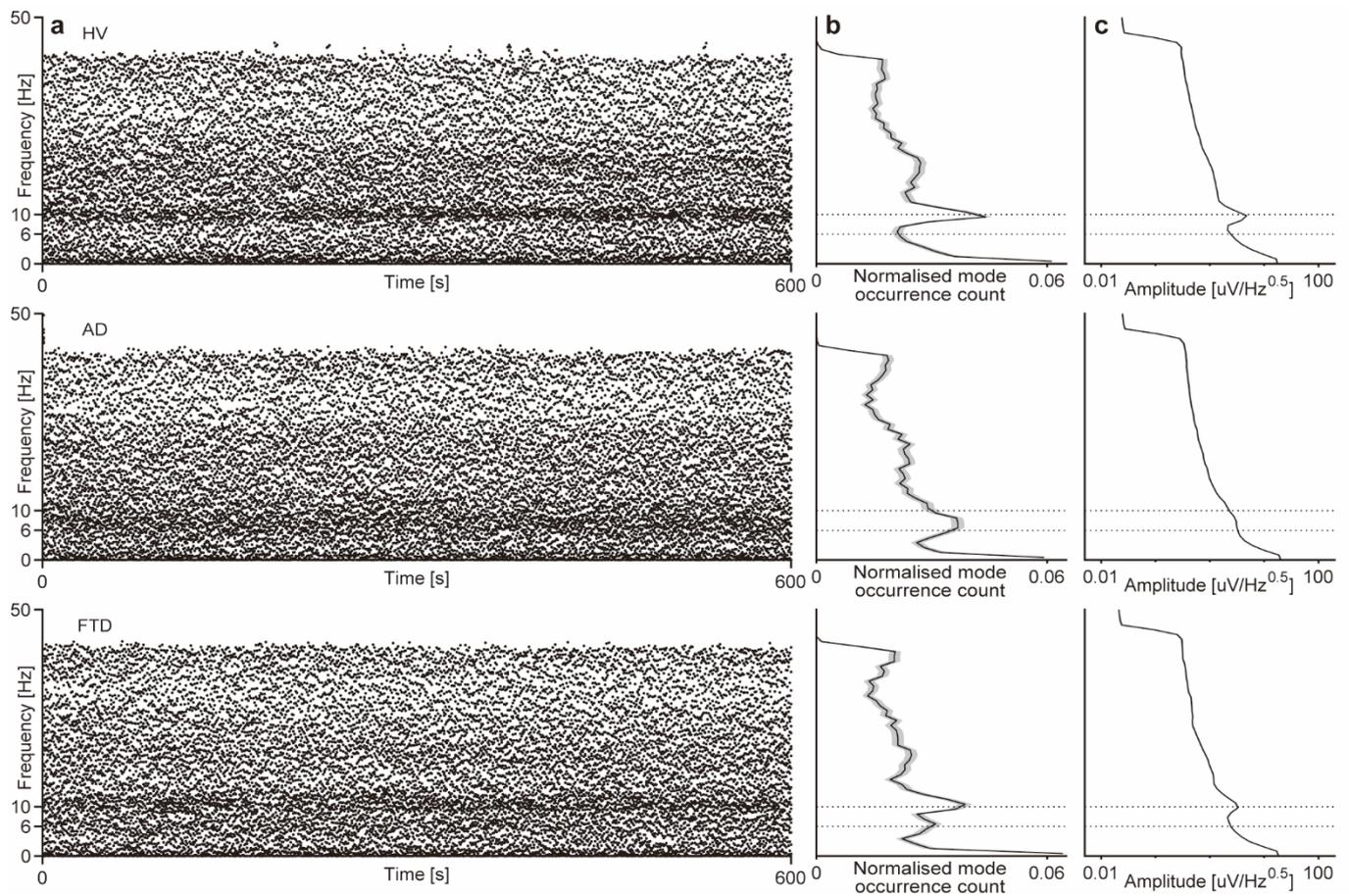

**Fig. 2. Frequency distribution of DMs and amplitude spectrum for representative participants.** (**a-c**) For a representative participant in each participant group of healthy subjects (HV), patients with Alzheimer's disease (AD), and patients with frontotemporal dementia (FTD), the following were visualised: (**a**) time series of the DM frequencies for each 1 s time window, (**b**) mean distribution of DM frequencies, and (**c**) mean amplitude spectrum. Visualisation in **a** was performed for only the first 600 seconds. In **b** and **c**, the 95% CI for the 1 s time window is shown as a shaded area. For the visualisations in **a** and **b**, the number of SVD components used in DMD was fixed at 50, which maximised the classification accuracy for the three groups when tfDM features were used.

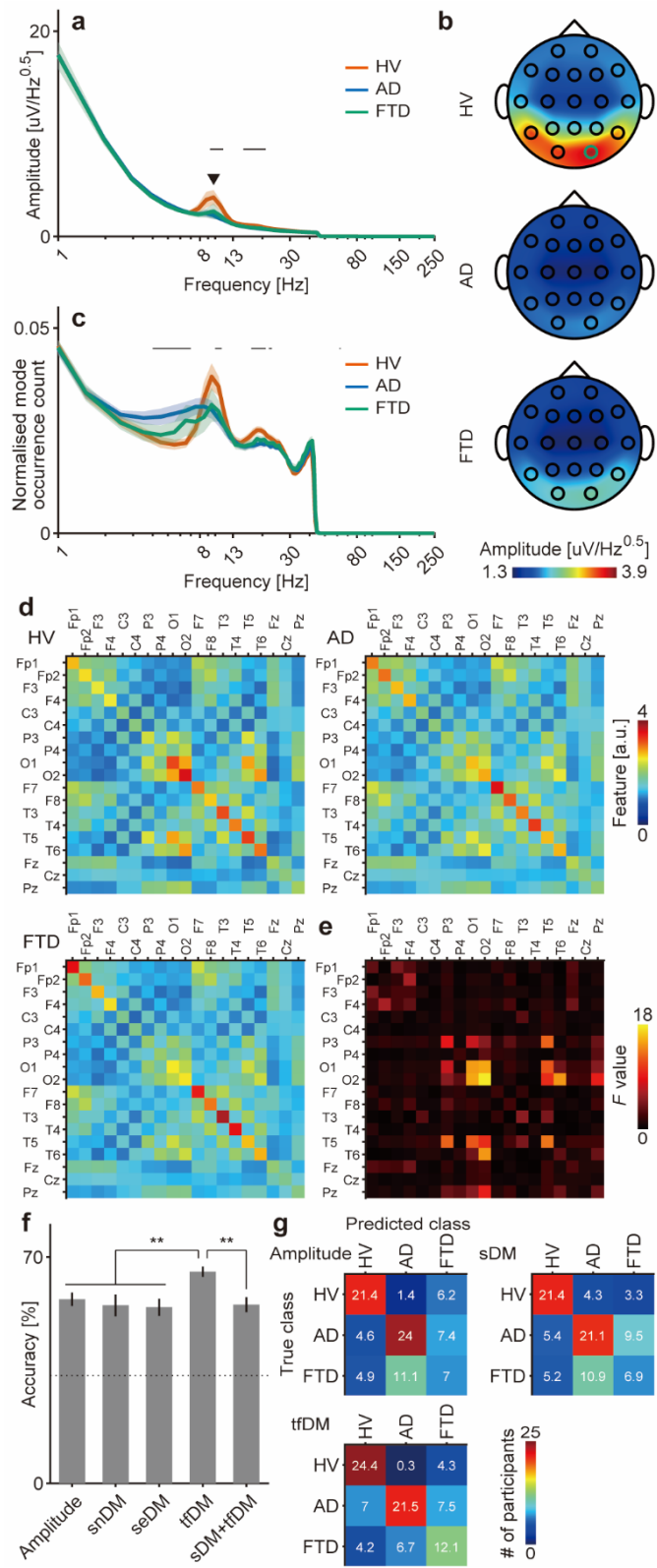

**Fig. 3. EEG activity of patients with dementia.** (**a**) Mean amplitude spectrum among healthy subjects, patients with AD, and patients with FTD for a representative channel

is shown with corresponding 95% CIs for shaded areas. The channel used for visualisation is shown in green in **b**. Significant differences between the participant groups are shown with black bars (Bonferroni-adjusted $p < 0.05$, one-way ANOVA). (**b**) The topography of the alpha amplitude is shown with colour coding. The frequency of the plot is indicated by the black arrow in **a**. (**c**) The mean distributions of the DM frequencies are shown with corresponding 95% CIs in shaded areas. Significant differences among the groups are shown with black bars (Bonferroni-adjusted $p < 0.05$, one-way ANOVA). For visualisation, the number of SVD components used in DMD was fixed at 50, which maximised the classification accuracy of the three groups using tfDM features. (**d**) Mean sDM features for each group and (**e**) the $F$ values among the groups are visualised with colour coding. (**f**) Classification accuracies of the three groups using different features are shown with bars. The error bar denotes the 95% CI among 10 repetitions of the classification. *$p < 0.05$, **$p < 0.01$, one-way ANOVA with a post hoc Tukey–Kramer test. (**g**) The confusion matrix of the classification shown in **f** is shown with colour coding for the amplitude, sDM, and tDM features.

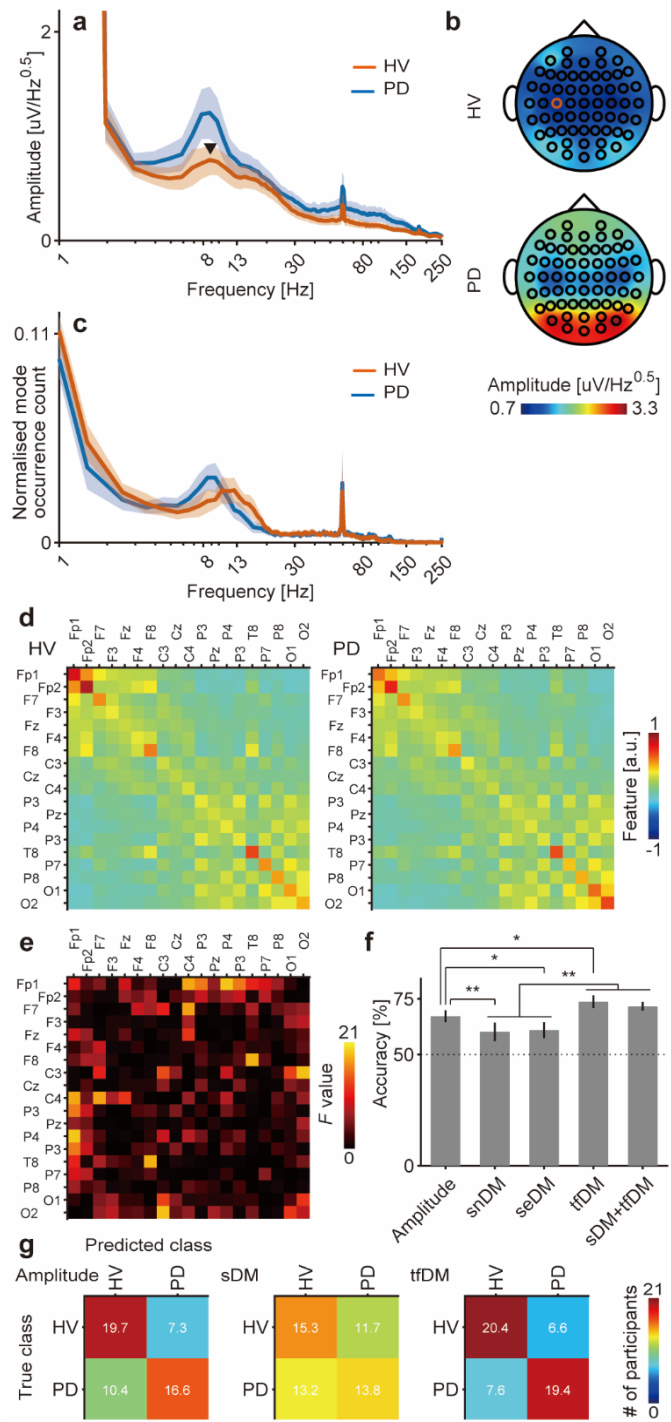

**Fig. 4. EEG activity of patients with Parkinson's disease.** (**a**) Mean amplitude spectrum among healthy subjects and patients with Parkinson's disease (PD) for a representative channel is shown with corresponding 95% CIs in shaded areas. The channel used for visualisation is shown in green in **b**. There were no significant differences between the participant groups (Bonferroni-adjusted $p >= 0.05$, one-way

ANOVA). (**b**) The topography of alpha amplitude is shown with colour coding. The frequency of the plot is indicated by the black arrow in **a**. (**c**) The mean distributions of the DM frequencies are shown with corresponding 95% CIs in shaded areas. Significant differences between the groups are only observed at 0–1 Hz, which is not shown in the plot (Bonferroni-adjusted $p < 0.05$, one-way ANOVA). For visualisation, the number of SVD components used in DMD was fixed at 20, which maximised the classification accuracy of the two groups using tfDM features. (**d**) Mean sDM features for each group and (**e**) the $F$ values among the groups are visualised with colour coding. (**f**) Classification accuracies of the two groups using different features are shown with bars. The error bar denotes the 95% CI among 10 repetitions of the classification. *$p < 0.05$, **$p < 0.01$, one-way ANOVA with a post hoc Tukey–Kramer test. (**g**) The confusion matrix of the classification shown in **f** is shown with colour coding for the amplitude, sDM, and tDM features.

**Method**
**Datasets**
In this study, two open EEG datasets (dementia[30] and PD[33]) and one dataset generated for the simulation analysis (Fig. 1) were used.

***EEG dataset for patients with dementia (open dataset from Miltiadous et al., 2023[30])***
*Dataset overview*
The studied dataset consists of EEG signals from 36 patients with AD, 23 patients with FTD, and 29 age-matched healthy subjects. Data recording was performed while the participants closed their eyes in the resting state. EEG signals were recorded at a sampling frequency of 500 Hz with electrode placement according to the 10–20 international system. The mean durations of the EEG signals for patients with AD, patients with FTD, and healthy subjects were 13.5, 12, and 13.8 min, respectively.

*Signal preprocessing*
In this study, pre-processed EEG signals in the /derivatives directory were analysed. Notably, pre-processing was performed via the following procedure: (1) filtering via a Butterworth bandpass filter (0.5 to 45 Hz), (2) re-referencing to average A1 and A2, (3) artefact subspace reconstruction, and (4) artefact rejection via independent component analysis.

***EEG dataset for patients with PD (open dataset from Cavanagh et al., 2018[33])***
*Dataset overview*
This dataset consists of resting-state EEG signals from 28 patients with PD and 28 healthy subjects. Event information was missing for one patient (803), and the signal duration was too short for one healthy subject (8070); hence, after removing the data for two participants, the data for 27 participants from both the healthy and patient groups were analysed. Data recording was performed for 1 min when the participants closed their eyes and when the patients were medicated. The EEG signals were recorded with a sampling rate of 500 Hz and CPz reference using a 64-channel brain vision system.

*Signal preprocessing*
Following the access code distributed with the dataset, four noisy channels were removed prior to further analysis.

***Simulation dataset***

*Dataset overview*

This dataset was generated to evaluate the behaviour of the DM frequencies when the signal contains time-varying components. This dataset consists of 100 trials of spatiotemporal signals for each stationary and nonstationary condition. The spatiotemporal signal of each trial spans 10 channels and has a duration of 1 s at a sampling rate of 500 Hz.

*Signal generation*

The spatiotemporal signal for the *j*-th trial was generated via the following equations.
Nonstationary condition:

$$\begin{cases} x_i^j(t) := 2 \cdot f^j(t) \cdot \sin(2\pi \cdot 10 \cdot t + \varepsilon_p) + 1 \cdot \sin(2\pi \cdot 20 \cdot t + \varepsilon_p) + 2.5 \cdot N(0,1) \text{ where } i \in \{1,2\} \\ x_i^j(t) := 2.5 \cdot N(0,1) \text{ where } i \in \{3, \cdots, 10\} \end{cases}$$

Stationary condition:

$$\begin{cases} x_i^j(t) := 1 \cdot \sin(2\pi \cdot 10 \cdot t + \varepsilon_p) + 1 \cdot \sin(2\pi \cdot 20 \cdot t + \varepsilon_p) + 2.5 \cdot N(0,1) \text{ where } i \in \{1,2\} \\ x_i^j(t) := 2.5 \cdot N(0,1) \text{ where } i \in \{3, \cdots, 10\} \end{cases}$$

In the equations, *i* and *t* denote channels ($i \in \{1, \cdots, 10\}$) and time ($t = 0, 0.02, \cdots, 1$), respectively, where $N(0,1)$ denotes the standard normal distribution. The initial phase of the sinusoidal wave ($\varepsilon_p$) was always determined randomly ($\varepsilon_p \sim U(-\pi, \pi)$, where $U$ denotes a continuous uniform distribution). $f^j(t)$ controls the on/off state of the 10 Hz signal component ($f^j(t) \in \{0,1\}$). For half of the nonstationary condition trials (50 trials), the 10 Hz signal component included the transition from off to on ($f_{off \to on}(t)$) at random times between 0.1 and 0.9 s when generating the signal. For the remaining trials, the signal included the transition from on to off ($f_{on \to off}(t)$) using the same transition time.

$$\begin{cases} f_{off \to on}^j(t) = (1 + sign(\sin(\pi t + \varepsilon_t^j)))/2, \text{ where } \varepsilon_t^j \sim U(-0.9\pi, -0.1\pi) \text{ and } j \in \{1, \cdots, 50\} \\ f_{on \to off}^j(t) = (1 + sign(\sin(\pi t + \varepsilon_t^j)))/2, \text{ where } \varepsilon_t^j = \varepsilon_t^{j-50} \text{ and } j \in \{51, \cdots, 100\} \end{cases}$$

**DMD**

Assuming that spatiotemporal signals are generated by a single dynamic system, the system can be described as follows:

$$\frac{d\mathbf{x}}{dt} = f(\mathbf{x}, t; \mu) \quad \ldots (1)$$

where $f(\cdot)$ denotes the dynamics with system parameter $\mu$ and $X(t) \in \mathbb{R}^P$ is a vector representing the state of the dynamic system measured at $P$ observation points at time $t$. Since the actual measurement of the signals is performed at discrete times with an interval of $\Delta t$, the dynamic system in discrete-time form corresponding to Eq. (1) can be introduced as follows:

$$x_{l+1} = F(x_l)$$

where $x_l$ denotes the $l$-th measurement of the system ($x_l = x(l\Delta t); \; l = 1, 2, \ldots, L$). In general, it is impossible to obtain $F$ analytically, and in practice, $F$ needs to be estimated from measurements. Here, $F$ is linearly approximated as

$$\mathbf{x}_{l+1} = \mathbf{A}\mathbf{x}_l \ldots (2).$$

$\mathbf{A}$ can be obtained by minimising the approximation error $\|\mathbf{x}_{l+1} - \mathbf{A}\mathbf{x}_l\|_2$ across all measurements of $l = 1, 2, \ldots, L-1$. By introducing two matrices of measurements, $\mathbf{X}$ and $\mathbf{X}'$, the linear approximation in Eq. (2) can be written as

$$\mathbf{X}' \approx \mathbf{A}\mathbf{X} \ldots (3)$$

where

$$\mathbf{X} = [\mathbf{x}_1 \ldots \mathbf{x}_{L-1}],$$
$$\mathbf{X}' = [\mathbf{x}_2 \ldots \mathbf{x}_L].$$

Notably, since DMD was developed for fluid dynamics analysis, it was assumed that the number of observation points ($P$) for $X$ was greater than the number of time points ($L$) ($P \gg L$). In this study, this condition is satisfied by performing signal stacking (see *Signal stacking*). The optimised $\mathbf{A}$ in Eq. (3) is obtained via the Moore–Penrose pseudoinverse (+) as $\mathbf{A} = \mathbf{X}'\mathbf{X}^+$ …(4). By applying SVD to $\mathbf{X}$, $\mathbf{X}$ can be decomposed to

$$\mathbf{X} \approx \mathbf{U}\Sigma\mathbf{V}^*$$

where $\mathbf{U} \in \mathbb{C}^{P \times K}$, $\Sigma \in \mathbb{C}^{K \times K}$, $\mathbf{V} \in \mathbb{C}^{L \times K}$, * represents the conjugate transpose, and $K$ denotes the number of SVD components used in the SVD approximation. Here, the left and right singular matrices ($\mathbf{U}$ and $\mathbf{V}$, respectively) are unitary, i.e., $\mathbf{U}^*\mathbf{U} = \mathbf{I}$ and $\mathbf{V}^*\mathbf{V} = \mathbf{I}$, and $\Sigma$ is a diagonal matrix composed of singular values. Notably, this SVD approximation assumes a low-dimensional structure of the dynamics. Via this decomposition, the pseudoinverse can be obtained as

$$\mathbf{X}^+ = \mathbf{V}\Sigma^{-1}\mathbf{U}^* \ldots (5)$$

From Eqs. (4) and (5), $\mathbf{A}$ can be obtained as follows:

$$\mathbf{A} = \mathbf{X}'\mathbf{V}\Sigma^{-1}\mathbf{U}^*$$

The eigenvector of $\mathbf{A}$ corresponds to DMs; however, the eigenvalue decomposition of $\mathbf{A}$ requires considerable computational resources because the number of observation points ($P$) is large (as mentioned above, $P \gg L$ is assumed). Considering that the rank of A is at most $L-1$, the DMs are obtained through eigen-decomposition of the low-dimensional matrix $\tilde{\mathbf{A}}$, which can be represented as follows via the orthogonal matrix $\mathbf{U}$:

$$\tilde{\mathbf{A}} = \mathbf{U}^*\mathbf{A}\mathbf{U} = \mathbf{U}^*\mathbf{X}'\mathbf{V}\Sigma^{-1}$$

By applying eigendecomposition to $\tilde{\mathbf{A}}$, sets of eigenvectors and eigenvalues are obtained as follows:

$$\tilde{\mathbf{A}}\mathbf{W} = \mathbf{W}\Lambda$$

where each column in $\mathbf{W}$ is an eigenvector and $\mathbf{\Lambda}$ is the diagonal matrix of the corresponding eigenvalues $\lambda_k$. The eigenvectors of $\mathbf{A}$ (DMs) are reconstructed from $\mathbf{W}$ and $\mathbf{\Lambda}$. Additionally, the eigenvalues and eigenvectors of $\mathbf{A}$ are given by $\mathbf{\Lambda}$ and the columns in $\mathbf{\Phi}$ ($\varphi$), respectively:

$$\mathbf{\Phi} = \mathbf{X'V\Sigma^{-1}W}.$$

With the introduction of the variable $\omega_k = ln(\lambda_k)/\Delta t$, the dynamics of $X$ can be approximated as follows:

$$\mathbf{x}(t) \approx \sum_{k=1}^{K} \boldsymbol{\varphi}_k e^{\omega_k t} b_k \quad \ldots (6)$$

Here, $\mathbf{b} = (b_1, \cdots, b_K)^T$ are the initial conditions of the modes and can be obtained as $\mathbf{b} = \mathbf{\Phi}^+ \mathbf{x}(0)$. By rewriting $\omega_k$ in Eq. (6), this expression can be transformed into a form that explicitly encompasses the decay/growth rate and frequency.

$$\mathbf{x}(t) \approx \sum_{k=1}^{K} \boldsymbol{\varphi}_k r_k^t \exp(2\pi i f_k t) b_k,$$

where $r_k = |\lambda_k|^{1/\Delta t}$ and $f_k = \arg(\lambda_k)/2\pi\Delta t$.

**Signal stacking**

DMD was originally developed for fluid dynamics analysis; hence, it is assumed that the number of observation points ($P$) for $X$ is greater than the number of time points ($L$) ($P \gg L$). However, for electrophysiological signals, $P$ is often smaller than $L$. In this case, the signals $\mathbf{X}$ and $\mathbf{X'}$ can be stacked $h$ times along the dimension of $P$ to fulfil the $P > L$ condition as follows:

$$\mathbf{X} = \begin{bmatrix} \mathbf{x}_1 & \mathbf{x}_2 & \cdots & \mathbf{x}_{L-h} \\ \mathbf{x}_2 & \mathbf{x}_3 & \cdots & \mathbf{x}_{L-h+1} \\ \vdots & \vdots & \ddots & \vdots \\ \mathbf{x}_h & \mathbf{x}_{h+1} & \cdots & \mathbf{x}_{L-1} \end{bmatrix},$$

$$\mathbf{X'} = \begin{bmatrix} \mathbf{x}_2 & \mathbf{x}_3 & \cdots & \mathbf{x}_{L-h+1} \\ \mathbf{x}_3 & \mathbf{x}_4 & \cdots & \mathbf{x}_{L-h+2} \\ \vdots & \vdots & \ddots & \vdots \\ \mathbf{x}_{h+1} & \mathbf{x}_{h+2} & \cdots & \mathbf{x}_L \end{bmatrix}.$$

In this study, $h$ was selected as the minimum integer that satisfies $h \geq \frac{L+1}{P+1}$. Since this stacking process results in $hP$ DMs, the first $P$ DMs and corresponding $\lambda$ are analysed.

**Acquisition of DMs and conversion to sDM/tfDM features**

For each dataset, the signals were divided on the basis of nonoverlapping 1 s time windows (500 samples) for DMD. To visualise the distribution of the DM frequencies (Figs. 2b, 3c, 4c), bins of 0–250 Hz divided by 1 Hz were created, and the number of mode occurrences within each bin was counted. To convert the DMs to sDM features, each DM ($\boldsymbol{\varphi}_k$) was L2-noramlised following the method in our previous study[27] to acquire sDM features as $\mathbf{\Phi\Phi}^\dagger$. Additionally, tfDM features were obtained by counting the number of DMs with frequencies that fell within a given frequency bin; here, the following conventional frequency bins were used: 0–1, 1–4, 4–8, 8–13, 13–30, 30–80, and 80–250 Hz. Finally, the sDM and tfDM features were averaged among the time windows to obtain features corresponding to each participant.

**Acquisition of amplitude spectra and amplitude features**

Amplitude spectra and amplitude features were calculated from the same 1-s signals that were used to calculate the DMs and DMD features. For each channel, the amplitude spectrum was acquired via a Hamming window and fast Fourier transform (FFT) of 512 points. To obtain the amplitude features of each subject, the amplitude spectrum of each channel was averaged within each of the same frequency bins used to calculate the tfDM features (0–1, 1–4, 4–8, 8–13, 13–30, 30–80, 80–250 Hz) and was averaged across the time windows.

**Classification analysis**

*Nested cross-validation*

To avoid over-estimation of the classification accuracy, hyperparameters used to train a classifier (cost parameter of the classifier and number of SVD components used in DMD) need to be optimised independently from the testing samples (i.e., for only the training samples); hence, in this study, 10-fold nested cross-validation was introduced for all classifications. For each outer cross-validation, the optimised parameter was estimated based on inner cross-validation within the training samples (from the outer fold); a classifier was then trained using all the training samples and the optimised parameter to classify the testing samples from the outer fold. The assignment of each subject to each fold was carried out by dividing each subject group into 10 groups in a way that minimised the variation in the number of subjects. To accurately estimate the classification accuracy, the assignment process was repeated 10 times randomly, and subsequent classification was performed. The assignment of subjects kept the same when different features were used for classification.

*Classifier training*

Classification was performed with the L1-regularised SVM included in LIBLINEAR 1.8[44] with default parameters, except for the use of the L1-regularised logistic regression solver (s: 6) and cost parameter (c). The cost parameter and number of SVD components used in DMD to acquire DM features were optimised from $10^{-1}, 10^0, \cdots,$ and $10^8$ and $1, 2, \cdots, 10, 15, \cdots, 50, 100, \cdots, 450$, full components, respectively. In cases where the number of training samples for each class was unbalanced, the samples for the class with fewer samples were repeatedly included so that the number of samples for all classes was the same. Furthermore, the classification accuracy was evaluated on the basis of balanced accuracy.

**Statistical test**

Amplitude spectra and normalised mode occurrence counts were assessed with one-way ANOVA among healthy/patient groups for each frequency (Figs. 3a, c, 4a, c), and the acquired *p* values were adjusted based on the number of frequencies (256) for the amplitude spectra and the number of frequency bins (250) for the distribution of the DM frequencies.

Classification accuracies using amplitude features, snDM features, seDM features, tfDM features, and combinations of sDM and tfDM features were compared via one-way ANOVA with a post hoc Tukey–Kramer test (Figs. 3f, 4f).

**Data availability:** No data measurements were performed in this study; only publicly available datasets were used.

**Code availability:** The code used in this study will be made available at https://github.com/yanagisawa-lab upon acceptance of the manuscript for publication.

**Acknowledgement:** This research was conducted under Japan Science and Technology Agency Moonshot R&D (JPMJMS2012, TY) and AIP Acceleration Research (JPMJCR24U2, TY). Additional support came from Japan Agency for Medical Research and Development (JP24he0122038, JP24wm0625207, JP24wm0625517, TY) and Sumino Isamu Foundation (TY).

**Author contributions:**
Ryohei Fukuma: conceptualization, methodology, software, validation, formal analysis, investigation, visualization, writing—original draft, writing—review and editing. Yoshinobu Kawahara: writing—review, and editing. Okito Yamashita: writing—review and editing. Kei Majima: writing—review and editing. Haruhiko Kishima: resources, writing—review and editing. Takufumi Yanagisawa: conceptualization, methodology, resources, writing—original draft, writing—review and editing, supervision, project administration, funding acquisition.

**Declaration of interests:** Takufumi Yanagisawa, Ryohei Fukuma, and Yoshinobu Kawahara are inventors on a pending patent application (Japanese Patent Application No. 2023-19430, filing date: November 24, 2023) related to the method reported in this work. The applicant for this patent is the University of Osaka.

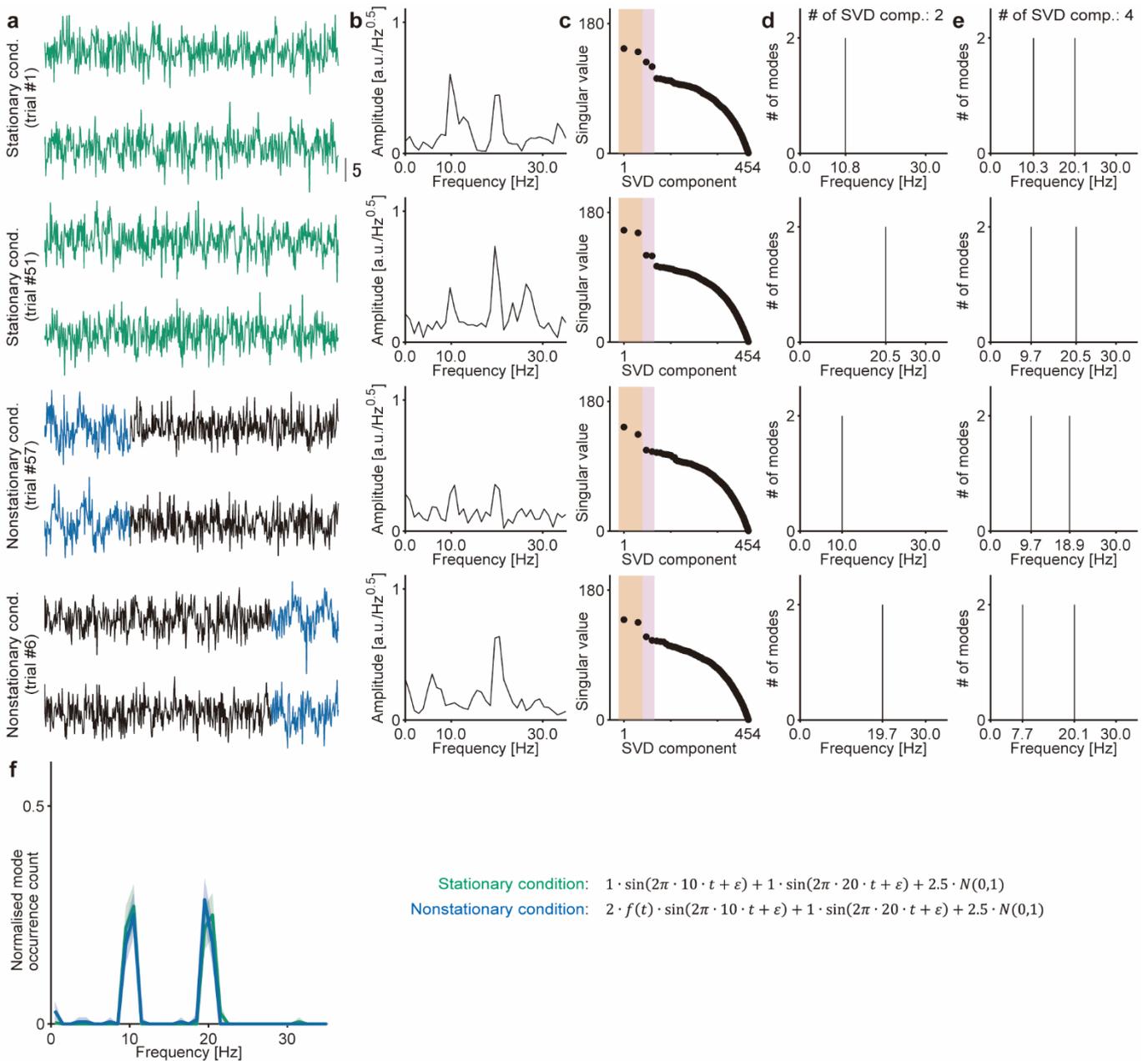

**Supplementary Fig. 1. The distribution of DM frequencies depends on the number of SVD components used in DMD.** (**a-e**) Representative trials for stationary and nonstationary conditions were visualised for (**a**) the analysed signal (1 s) in channels 1 and 2 (upper and lower lines, respectively), (**b**) the amplitude spectrum for the signal in channel 1, which is shown in **a**, (**c**) singular values, (**d**) the distribution of DM frequencies based on the first two SVD components (shown with red shaded areas in **c**), and (**e**) the distribution of DM frequencies based on the first four components (shown with red and pink shaded areas in **c**). The blue line in **a** denotes the interval that contains

the 10 Hz component. Note that trial #51 was used as a representative trial, as shown in Fig. 1a-g. (**f**) Distributions of the DM frequencies shown in Fig. 1h were recalculated on the basis of the first four SVD components during DMD and are shown in the same format as in Fig. 1h. The shaded areas represent the 95% CI among 100 trials.

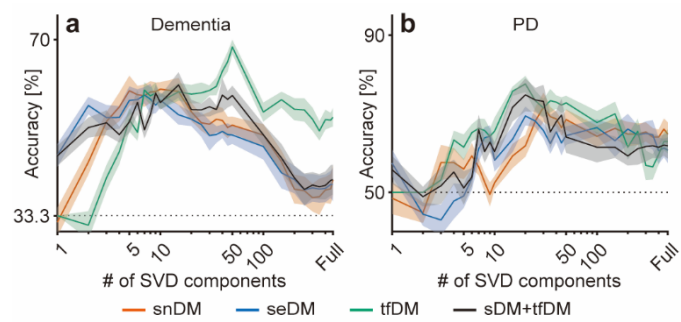

**Supplementary Fig. 2. Classification accuracies versus the number of SVD components used in DMD.** (**a**, **b**) Classification accuracies using different features derived from DMs were plotted against the number of SVD components used in DMD for (**a**) healthy subjects and patients with dementia and (**b**) healthy subjects and patients with PD. The shaded area denotes the 95% CI among 10 repetitions of the classification.

**Supplementary Fig. 3. All sDM features and $F$ values for patients with Parkinson's disease.** (**a**) Mean sDM features for each group and (**b**) the $F$ values among the groups visualised with colour coding.